\documentclass[preprint2]{aastex}

\usepackage{xcolor}

\newcommand{\kkmspc}{K km s$^{-1}$ pc$^2$}
\def\simgt{\lower.5ex\hbox{\gtsima}}
\def\simlt{\lower.5ex\hbox{\ltsima}}

\shortauthors{Tan et al.}

\begin{document}

\title{A deep search for molecular gas in two massive Lyman break galaxies at z=3 and 4: vanishing CO-emission due to low metallicity?}

\author{Q. Tan\altaffilmark{1,2,3}, E. Daddi\altaffilmark{2}, M. Sargent\altaffilmark{2}, G. Magdis\altaffilmark{4}, J. Hodge\altaffilmark{5}, M. B\'{e}thermin\altaffilmark{2}, F. Bournaud\altaffilmark{2}, C. Carilli\altaffilmark{6}, H. Dannerbauer\altaffilmark{7}, M. Dickinson\altaffilmark{8}, D. Elbaz\altaffilmark{2}, Y. Gao\altaffilmark{1}, G. Morrison\altaffilmark{9,10}, F. Owen\altaffilmark{6}, M. Pannella\altaffilmark{2}, D. Riechers\altaffilmark{11}, and F. Walter\altaffilmark{5}}

\email{qhtan@pmo.ac.cn}

\altaffiltext{1}{Purple Mountain Observatory \& Key Laboratory for Radio Astronomy, Chinese Academy of Science, Nanjing 210008, China}
\altaffiltext{2}{CEA Saclay, DSM/Irfu/Service d'Astrophysique, Orme des Merisiers, 91191 Gif-sur-Yvette Cedex, France}
\altaffiltext{3}{Graduate University of the Chinese Academy of Sciences, 19A Yuquan Road, Shijingshan District, Beijing 100049, China}
\altaffiltext{4}{Department of Physics, University of Oxford, Keble Road, Oxford OX1 3RH, UK}
\altaffiltext{5}{Max-Planck Institute for Astronomy, K\"{o}nigstuhl 17, 69117 Heidelberg, Germany}
\altaffiltext{6}{National Radio Astronomy Observatory, P. O. Box O, Socorro, NM 87801, USA}
\altaffiltext{7}{Universit\"{a}t Wien, Institut f\"{u}r Astronomie T\"{u}rkenschanzstrasse, 171160 Wien, Austria}
\altaffiltext{8}{National Optical Astronomical Observatory, 950 North Cherry Avenue, Tucson, AZ 85719, USA}
\altaffiltext{9}{Institute for Astronomy, University of Hawaii, Honolulu, HI, 96822, USA}
\altaffiltext{10}{Canada-France-Hawaii Telescope, Kamuela, HI, 96743, USA}
\altaffiltext{11}{Department of Astronomy, Cornell
   University, 220 Space Sciences Building, Ithaca, NY 14853, USA}

\begin{abstract}
We present deep IRAM Plateau de Bure Interferometer (PdBI) observations, searching for CO-emission toward two massive, non-lensed Lyman break galaxies (LBGs) at $z=3.216$ and 4.058. 
With one low significance CO detection (3.5$\sigma$)  and one sensitive upper limit, we find that the CO lines are $\gtrsim$3--4 times weaker than expected based on the relation between IR and CO luminosities followed by similarly, massive galaxies at $z=$0--2.5. This is consistent with a scenario in which these galaxies have low metallicity, causing an increased CO-to-H$_2$ conversion factor, i.e., weaker CO-emission for a given molecular (H$_2$) mass. The required metallicities at $z>3$ are lower than predicted by the fundamental metallicity relation (FMR) at these redshifts, consistent with independent evidence. Unless our galaxies are atypical in this respect, detecting molecular gas in normal galaxies at $z>3$ may thus remain challenging even with ALMA. 
\end{abstract}


\section{Introduction}

The study of cold gas, the fuel for star formation, in massive galaxies across cosmic time is crucial to understanding how galaxies have converted their gas into stars \citep[e.g., review by][]{carilli13}. Thanks to sensitivity improvements of millimeter interferometers, measurements of molecular gas have recently become feasible in normal\footnote{non-starbursting} galaxies at $z\sim$0.5--2.5 \citep{daddi08,daddi10,tacconi10,tacconi13,geach11}, which lie on the star-forming main-sequence (MS). Galaxies following this relation between stellar mass and star formation rate (SFR) contribute $\sim$90\% to the cosmic SFR density at these redshifts \citep{daddi07,rodighiero11,sargent12}. It is found that these high-z star-forming galaxies are rich in molecular gas, with gas fractions rising from $\sim$5\% locally to $\sim$50\% at $z\sim$2 for objects with $M_{\star}\sim10^{11}M_\odot$, an evolution similar to that of their specific star formation rates (sSFRs) \citep{magdis12b,tacconi13}. 

However, little is known about the gas content of normal galaxies at $z\gtrsim$3. Although CO detections have been reported in lensed Lyman break galaxies (LBGs) at $\gtrsim$3 \citep{riechers10,livermore12}, these lenses have very low stellar masses $(M_{\star}\lesssim10^9M_\odot$) and high sSFR, and thus might not be directly comparable to more massive main-sequence  galaxies at those epochs. The dependence of the CO-to-H$_2$ conversion factor on metallicity \citep[e.g.,][and references therein]{bolatto13,carilli13} may imply that CO-emission becomes weak in typical massive galaxies with low metal-enrichment at $z\gtrsim$3 \citep{tacconi08,genzel12,narayanan12}.  

In this Letter we present new, deeper CO observations of the $z\sim3$ LBG M23 (for which \citealt{magdis12a} reported a $\sim\ 4\sigma$ CO signal), as well as of the UV-selected LBG BD29079 at $z\sim4$ \citep{daddi09}. These two massive LBGs lie in the GOODS-North field. With the available optical and infrared data, we investigate their molecular gas properties and place constraints on the evolution of the molecular gas fraction, in the context of our current understanding of the metallicity-evolution of galaxies at those epochs. We assume $H_{\rm 0}=71$~km~s$^{-1}$ Mpc$^{-1}$, $\Omega_{\rm M}=0.3$,  $\Omega_\Lambda=0.7$.

\section{Observations}

We used the IRAM/PdBI to observe CO(4-3) emission from BD29079 and CO(3-2) from M23 (Table~\ref{tbl-1}). BD29079 is in the vicinity of GN20 (separation of $\sim16''$; \citealt{pope05}) and at the same redshift \citep{daddi09}. BD29079 was observed in three configurations. The pointing center of the AB-configuration observations lay $30''$ north of BD29079 \citep{daddi09}, while the new D-~and~C-configuration observations were centered on BD29079 and carried out under good 3mm weather conditions in June 2009 and January-April 2013. In total, the six antenna-equivalent on-source time for BD29079 is 13.8hr. For M23, new C-configuration observations were made in April 2013. The combination of C-~and~D-configuration \citep{magdis12a} observations give a total on-source time of 7.3hr. For observations made before and after 2011, correlator bandwidths are 1 GHz and 3.6 GHz (WideX), respectively. All observations were performed in dual polarization mode.

We reduced the data with the GILDAS software packages CLIC and MAPPING. After correcting for primary beam attenuation (PBA), the rms noise in a 26~km~s$^{-1}$ channel and at an angular resolution of $\sim3.0''$ and $\sim4.5''$ for BD29079 and M23 are 0.38~and~0.72~mJy/beam, respectively. 

For BD29079, we also make use of CO(6-5) and CO(2-1) data from deep PdBI and JVLA observations of the GN20 field originally presented in \citet{carilli10} and \citet{hodge12,hodge13}. To avoid over-resolving of the source, the CO(2-1) data cube was tapered to a spatial resolution of 1.4$''$ and binned to 26~km~s$^{-1}$ spectral resolution, with a corresponding noise of 0.21 mJy/beam. The sensitivity reached in the CO(6-5) data cube ($0.9''\times26$~km~s$^{-1}$) is 0.58~mJy/beam.

\begin{deluxetable}{lclllccllcc}
\tabletypesize{\scriptsize}
\small\addtolength{\tabcolsep}{-4pt}
\tablecaption{Summary of PdBI Observations\label{tbl-1}}
\tablewidth{0pt}
\tablehead{
\colhead{Source} & \colhead{R.A.} & \colhead{Decl.} & \colhead{Conf.} & \colhead{Obs. Dates} & \colhead{PBA} & 
\colhead{$T_{\rm int}\tablenotemark{a}$} & \colhead{Frequency} & \colhead{Combined Beam} &
\colhead{rms}$\tablenotemark{b}$ & \colhead{Reference}\\
\colhead{} & \colhead{(J2000)} & \colhead{(J2000)} & \colhead{} & \colhead{} &  \colhead{} &
\colhead{(hr)} & \colhead{(GHz)} & \colhead{} &
\colhead{($\mu$Jy)}  & \colhead{}

}
\startdata
BD29079 & 12:37:11.52 & 62:21:55.6 &  AB  &  2008 Jan-Feb & 2.4 &  2.2 & 91.375  & $2''.98\times2''.27$ P.A.=70$^\circ$ & 29  & \citet{daddi09}\\
                  &                        &                     &  D  &  2009 Jun          & 1.0 & 7.4 & 91.375   &    &  & This work \\
                  &                        &                     &  C  &  2013 Jan-Apr   & 1.0 & 4.2 & 91.375   &    &  & This work \\
\tableline
M23          & 12:37:02.70  & 62:14:26.3 & D  & 2011 May  & 1.0 & 2.5 & 82.059 & $4''.45\times3''.82$ P.A.=56$^\circ$ & 40 & \citet{magdis12a}\\
                  &                         &                     & C  & 2013 Apr    & 1.0 & 4.7 & 82.059 &   &  & This work \\
\enddata
\tablenotetext{a}{Six antenna-equivalent effective on-source integration time, corrected for PBA.}
\tablenotetext{b}{Noise per beam, averaged over full bandwidth.}
\end{deluxetable}

\section{Results}

\subsection{BD29079} \label{bd}

BD29079 is an UV-selected $B$-dropout galaxy (Figure~\ref{fig1}) with spectroscopic redshift $z=4.058$ \citep{daddi09}. CO measurements toward this galaxy were previously reported as upper limits \citep{carilli11,hodge13}. Combining our new, deep CO(4-3) data with those already published, we searched for CO-emission to deeper levels. Figure~\ref{fig1} shows CO spectra binned to 26~km~s$^{-1}$ at the expected frequency of CO~$J$=2-1, 4-3, and 6-5. No line emission from the individual CO transitions was detected. 
 
In order to maximize sensitivity, we performed a combined stack of the spectra covering all three CO transitions. The amplitude of all spectra is rescaled to the level appropriate for a template spectral line energy distribution (SLED). We consider scaling factors based on observed CO SLEDs of both sBzK galaxies and SMGs (Aravena et al. 2010; Carilli et al. 2010;  Daddi et al. 2013, in preparation). In the line stack each spectrum is weighted by the inverse square of the rescaled noise. Both CO-excitation models produce no detection of the stacked line emission and lead to a comparable flux limit. We adopt an upper 3$\sigma$ CO(1-0) luminosity limit of $<8.2\times10^9$ \kkmspc, derived from the flux limit of 4.1~mJy~km~s$^{-1}$ (1$\sigma$, assuming a line width of 300~km~s$^{-1}$, consistent with the typical line widths measured in similarly massive high-redshift star-forming galaxies; \citealp[see][]{daddi10,tacconi10}) obtained for the sBzK-excitation model\footnote{We adopt line temperature ratios of $r_{61}$=0.10, $r_{41}$=0.21, and $r_{21}$=0.70 (Aravena et al. 2010; Daddi et al. 2013, in preparation).}. Using SMG-like CO SLEDs or those of $z\gtrsim3$ lenses \citep{riechers10,coppin07,baker04} would imply even deeper upper limits.

\subsection{M23} \label{m23}
\begin{figure*}
\centering
\begin{minipage}[b]{0.65\linewidth}
\centering
\includegraphics[width=0.45\textwidth]{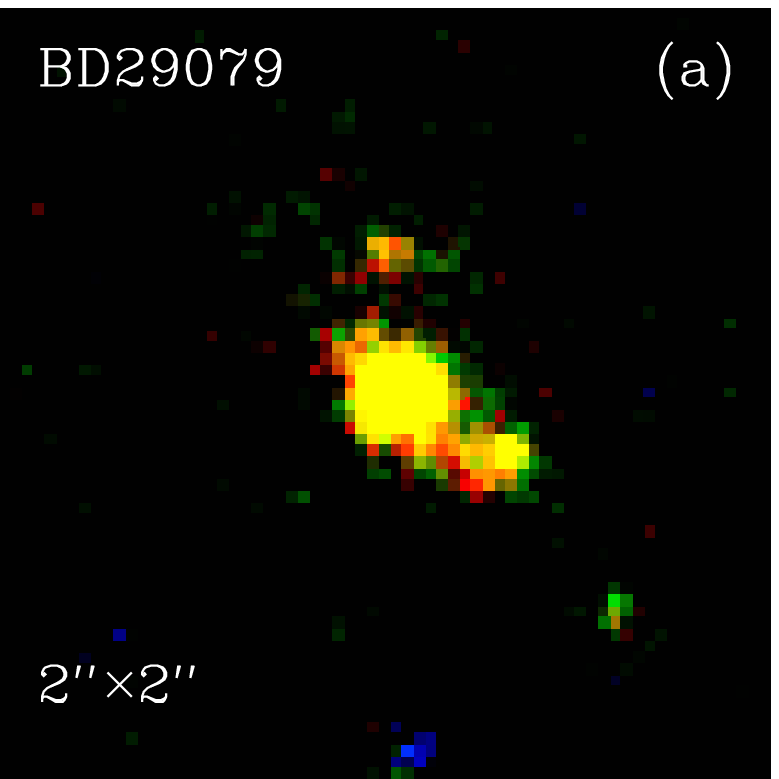}
\includegraphics[width=0.45\textwidth]{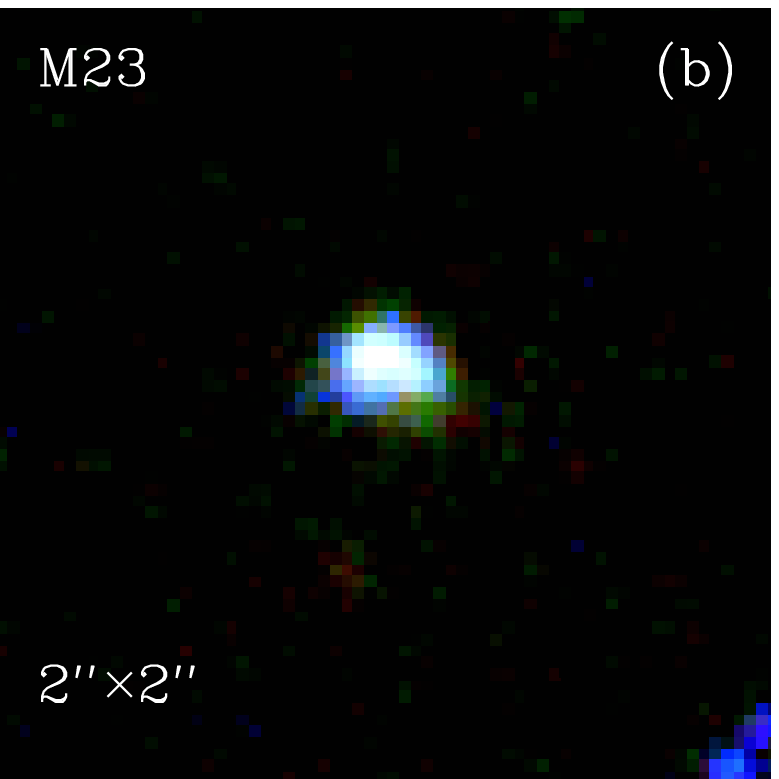}
\end{minipage}\\
\begin{minipage}[b]{0.65\linewidth}
\includegraphics[width=0.49\textwidth]{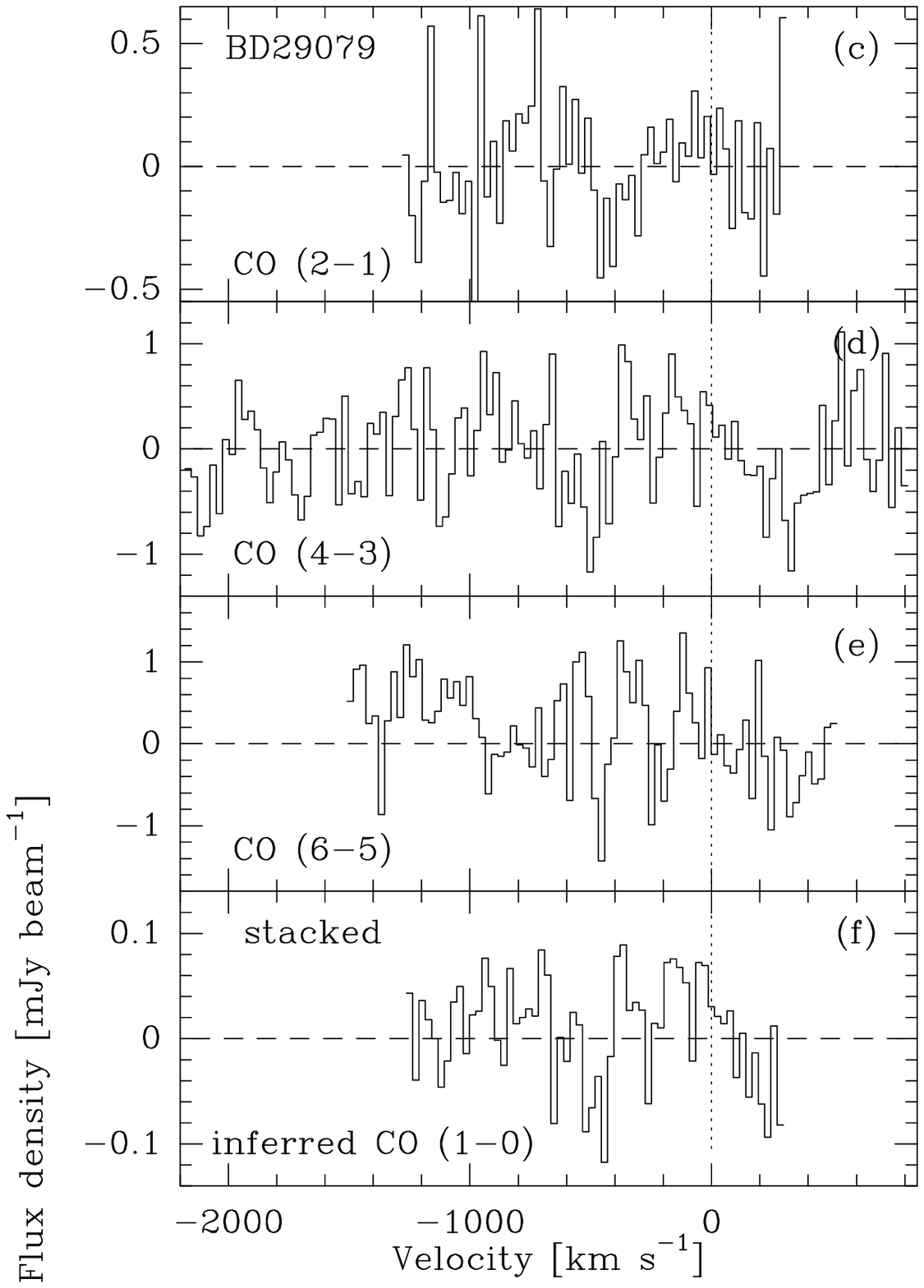}
\hspace{-0.25cm}
\includegraphics[width=0.49\textwidth]{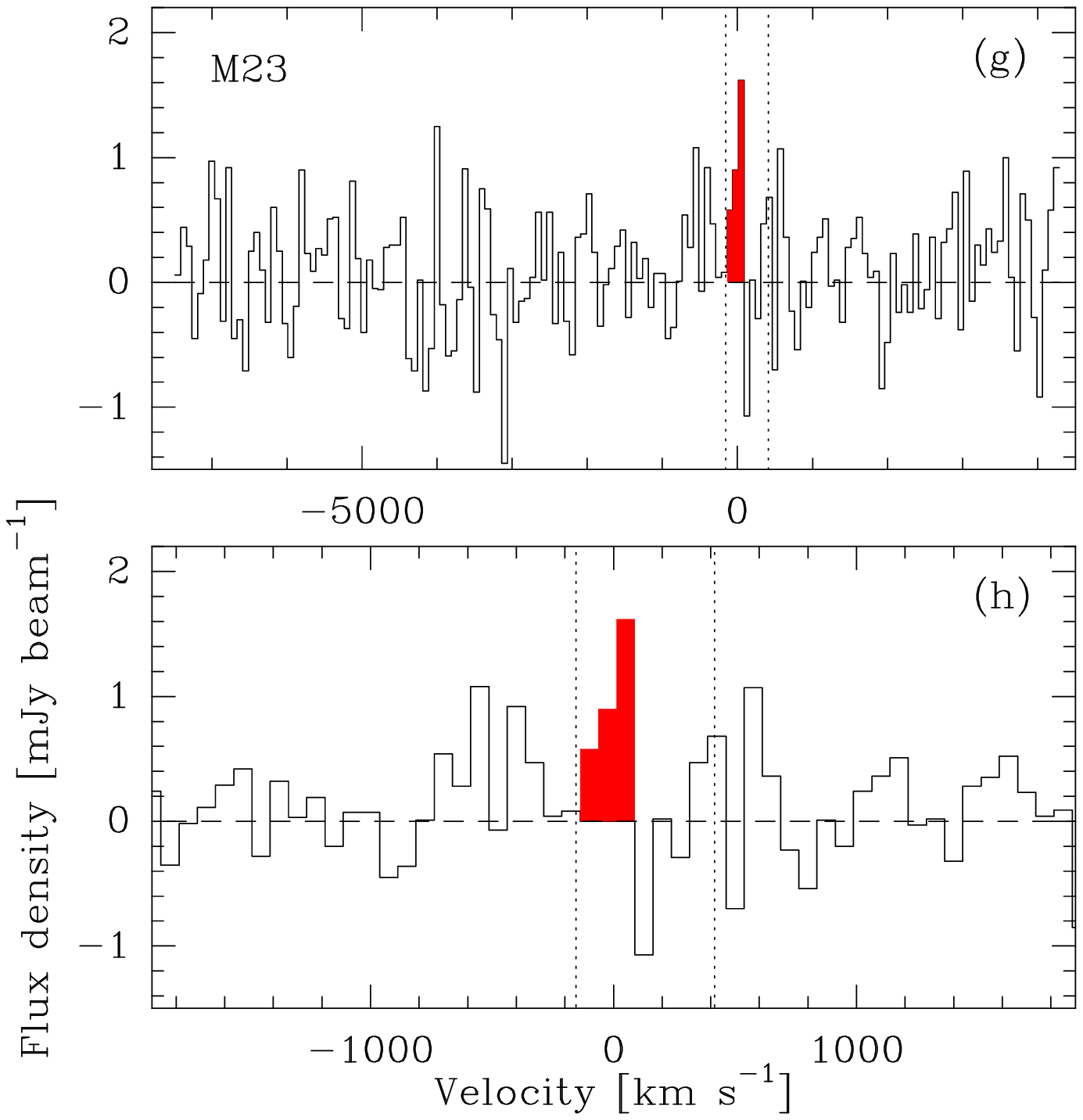}
\end{minipage}
\caption{\scriptsize {\it Top}: HST+ACS color image of BD29079 (a) and M23 (b) (F435W-blue, F775W-green, and F850LP-red). The yellow color of BD29079 reflects its selection as a B-dropout. {\it Bottom}: CO spectra for BD29070 and M23. For BD29079, the frequency coverage of CO J=2-1(c), 4-3(d), 6-5(e), and of the stacked spectrum (f) is shown. (Spectral binning: 26~km~s$^{-1}$; zero velocity corresponds to z=4.058). The stacked spectrum is produced by assuming a sBzK CO excitation model. For M23, we show the CO(3-2) spectrum with 75~km~s$^{-1}$ spectral resolution over the full velocity range observed (h) and a zoom-in on the tentative line position (g). Red color indicates where the maximum S/N ratio is found within the velocity range given by redshift uncertainties (1$\sigma$; dotted lines) in the optical spectroscopy \citep[see][]{magdis12a}\label{fig1}}
\end{figure*}

\citet{magdis12a} reported CO(3-2) line emission with S/N=4 toward M23. We achieved no significant line-detection in the new observations, such that, in the combined dataset (threefold integration time increase relative to \citealt{magdis12a}), the overall significance decreases to $\sim\ 3.5\sigma$ (Figure~\ref{fig1}), suggesting that  the flux density reported in \cite{magdis12a} was boosted by noise. Based on the deeper data, we still consider the line feature in M23 as tentative and revise our estimate of its CO(3-2) luminosity to $(9.5\pm2.7)\times10^9$ \kkmspc, though this may still represent an upper limit. Assuming a correction factor of $r_{31}=0.5$ (measured in high-z normal galaxies; \citealp[e.g.,][]{dannerbauer09,tacconi10}), the CO(1-0) luminosity can be determined from CO(3-2) and is listed in Table~\ref{tbl-2}.

\begin{deluxetable}{lcccccccc}
\centering
\tabletypesize{\scriptsize}
\small\addtolength{\tabcolsep}{-4pt}
\tablecaption{Observed and Derived Properties\label{tbl-2}}
\tablewidth{0pt}
\tablehead{
\colhead{Source} & \colhead{\it{z}$_{\rm Keck}$}  & \colhead{$L'_{\rm CO}$} & \colhead{$M_{\rm gas}\tablenotemark{a}$} & \colhead{$\alpha_{\rm CO}\tablenotemark{b}$} &
\colhead{SFE} & \colhead{$M_{\star}\tablenotemark{c}$} & \colhead{$L_{\rm IR}$} & \colhead{sSFR}  \\
\colhead{} & \colhead{}  & \colhead{(10$^{10}$~K~km~s$^{-1}$~pc$^2$)} & \colhead{($10^{10}\ M_{\odot}$)} & \colhead{($M_{\odot}$ (K~km~s$^{-1}$~pc$^2$)$^{-1}$)} & 
\colhead{($L_{\odot}$ (K~km~s$^{-1}$~pc$^{-2}$)$^{-1}$)} & \colhead{($10^{10}\ M_{\odot}$)} & \colhead{($10^{12}\ L_{\odot}$)} & \colhead{(Gyr$^{-1}$)} 

}
\startdata
BD29079 & 4.058 &   $<0.82\tablenotemark{d}$  &  $\sim$16 &  $>$19 & $>$330 & $2.5\ \pm0.5$ &  $2.7\ \pm \ 0.6$  & 10.8  \\
M23          & 3.216 &   1.90$\pm$0.54 & $\sim$18 & $\gtrsim$9.3 & $\gtrsim 160$ & $10\pm 5$ & $3.1\pm1.1$  & 3.1  \\
\enddata

\tablenotetext{a}{From the integrated Schmidt-Kennicutt relation, $M_{\rm gas}$ vs. SFR (see text).}
\tablenotetext{b}{$\alpha_{\rm CO}$=$M_{\rm gas}/L'_{\rm CO}$}
\tablenotetext{c}{Stellar mass from SED fitting, assuming a Chabrier IMF}
\tablenotetext{d}{3$\sigma$ upper limit on CO(1-0) assuming FWHM=300~km~s$^{-1}$. }
\end{deluxetable}

\section{Analysis and Discussion}
\subsection{Physical properties}\label{properties}

For M23, \citet{magdis12a} estimate SFR = (310$\pm$110) M$_\odot$ yr$^{-1}$, based on radio and 250$\mu$m detections. After accounting for flux from a nearby (likely unrelated, compact, red) companion source, we revise the stellar mass estimate for M23 to $M_{\star} = (1.0\pm0.5)\times 10^{11}$ M$_\odot$, implying a sSFR$\sim$ 3.1 Gyr$^{-1}$. This value is close to the average sSFR of equally massive galaxies on the SFR-$M_{\star}$ main-sequence at $z\sim$3 (sSFR/sSFR$_{\rm MS}\sim$1.4), suggesting that M23 is a normal galaxy rather than a starburst, notwithstanding its high SFR.

BD29079 has a stellar mass of $M_{\star} = (2.5\pm0.5) \times 10^{10} $ M$_\odot$, as determined by SED-fitting of the rest-frame UV$-$to$-$near-IR broad-band photometry using \citet{bruzual03} models and assuming a constant star formation history (SFH), a \citet{chabrier03} IMF, and a \citet{calzetti00} extinction law. We infer an extinction-corrected 1500\AA\ rest-frame luminosity corresponding to SFR of $170\pm40{\rm M}_\odot$~yr$^{-1}$(see Table~\ref{tbl-2}). Complementary, independent constraints on the $L_{\rm IR}$ of BD29079 follow from its JVLA and Herschel coverage. We detect a 14$\mu$Jy-signal (4$\sigma$) in the 1.4~GHz JVLA map of  Owen et al. (2013, in preparation), only slightly offset from the nominal HST position of the source. When combining this constraint with the multi-band Herschel non-detections with PACS and SPIRE, with a measurement of $0.2\pm0.1$mJy for the 2.2mm continuum and an upper limit at 3.3mm in our PdBI data, we derive a combined estimate of $L_{\rm IR}$ = $(2.7\pm0.6)\times10^{12}$ L$_\odot$, consistent within uncertainties with the extinction-corrected UV SFR estimate. This corresponds formally to a sSFR-excess of $\approx$2--3 for BD29079, with respect to the characteristic value of an average main-sequence galaxy with identical mass and redshift, implying in principle a non-negligible starburst-probability for this source \citep[e.g.,][henceforth S13a]{sargent13}. While we cannot rule out the presence of a merger-induced starburst event, the fact that this galaxy was initially UV-selected, and the reasonable agreement between UV- and radio/FIR-derived SFR constraints (similar to the correspondence found for normal, massive star-forming galaxies at z$\sim$2;  see Daddi et al. 2007) disfavor this possibility. 

\begin{figure*}[t]
\centering
\begin{minipage}[b]{0.43\linewidth}
\centering
\includegraphics[width=\textwidth]{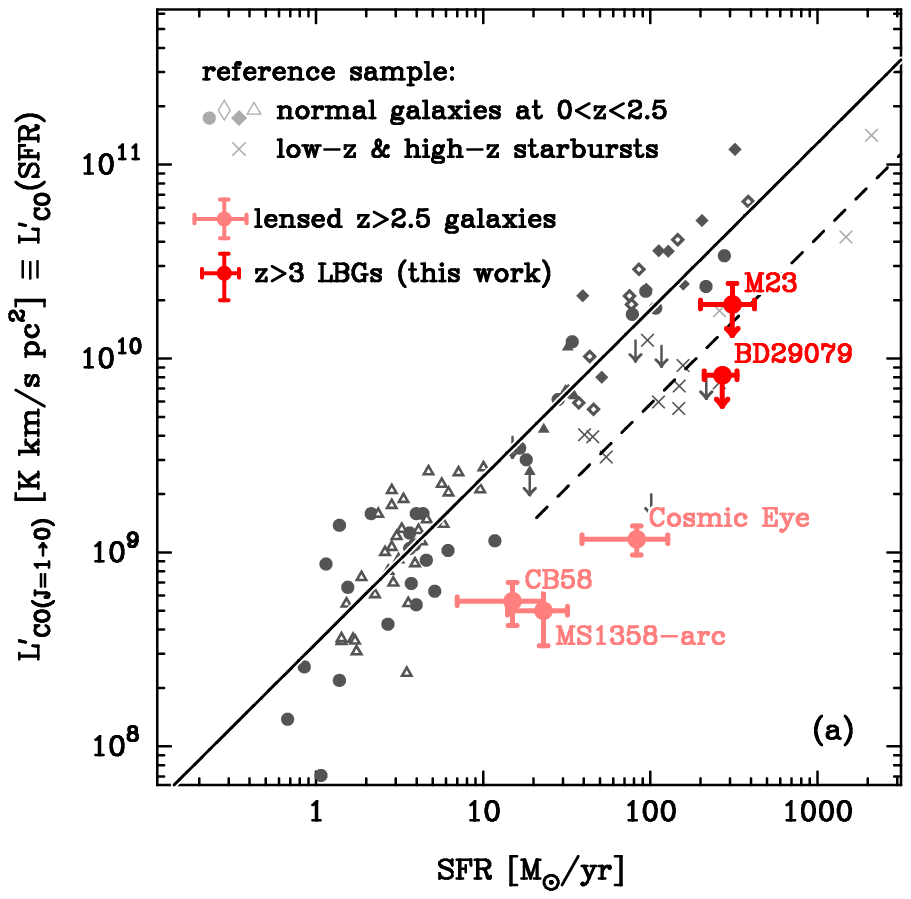}
\end{minipage}
\begin{minipage}[b]{0.43\linewidth}
\centering
\includegraphics[width=\textwidth]{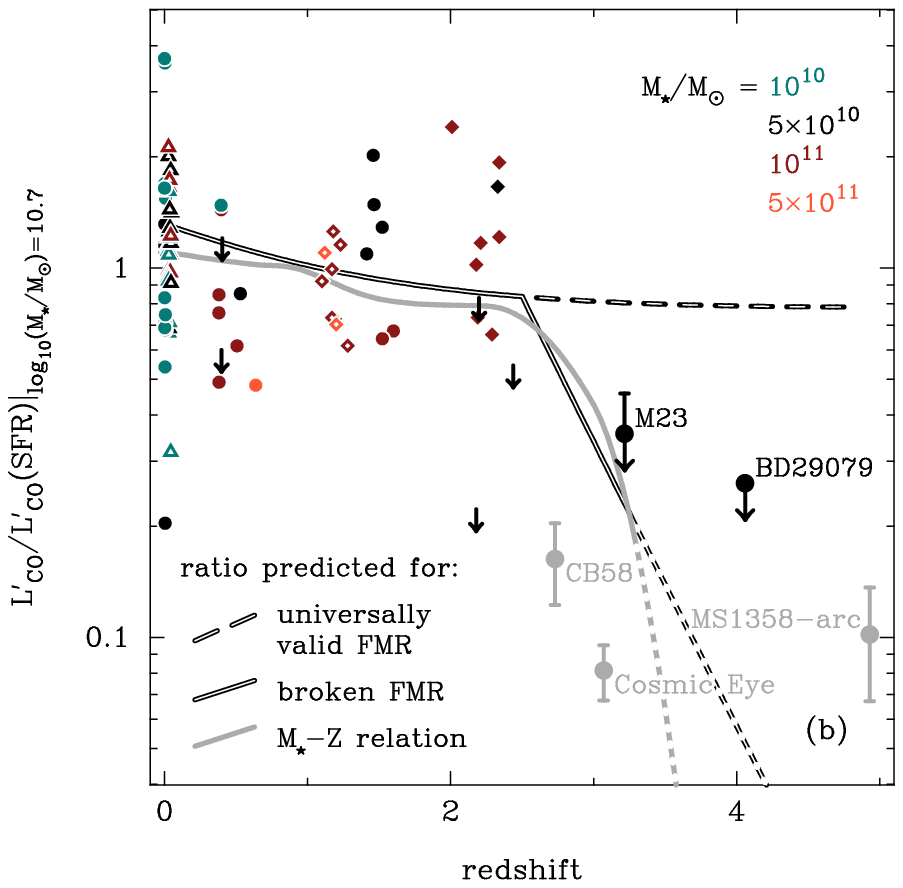}
\end{minipage}
\caption{\scriptsize {\it Left}: Location of non-lensed LBGs (bright-red dots) and lensed LBGs (pale-red dots) in SFR-L$'_{\rm CO}$ space. Filled/open grey symbols show measurements for normal galaxies and grey crosses represent starbursts at low- and high-redshift. 
See \citet{sargent13} for references to the literature data shown (different symbol shapes correspond to different samples).
 Solid and dashed lines--best-fitting relation derived for main-sequence galaxies and average offset of strong starbursts, respectively. {\it Right}: Redshift-evolution of the CO luminosity of galaxies, normalized by the average $L'_{\rm CO}$(SFR) trend for MS galaxies in panel (a). Predictions are shown for the case of an universally valid FMR/broken FMR/evolving $M_{\star}-Z$ relation at $z\geq$3 (dashed/solid/grey lines; the $z>3.5$ extrapolations are observationally-unconstrained). 
All measurements for normal galaxies (colored points) and predictions have been normalized to a common mass scale of $M_{\star}=5\times10^{10}M_\odot$ based on the expected mass dependence of $L'_{\rm CO}/L'_{\rm CO}({\rm SFR})\propto M_{\star}^{0.3}$. Data points are colored depending on which reference color was closest in stellar mass. 
No mass-dependent scaling was applied to the lensed, possibly starburst-like $z>3$ sources.
 \label{fig2}} 
\end{figure*}

\subsection{CO luminosity}\label{lco}

The ratio between the observables $L_{\rm IR}$ and $L'_{\rm CO}$ can be used as an indication of star formation efficiency (SFE). We find values of $\gtrsim 160$ and $>330 L_\odot$~(\kkmspc)$^{-1}$ for M23 and BD29079, respectively. These are $\gtrsim3$--4 times higher than the typical values defined by normal galaxies (Figure~\ref{fig2}a), which display a dispersion of 0.21~dex (S13a), implying that  both are 2--3$\sigma$ outliers.  We therefore find substantially weaker CO-emission in our two LBGs, compared to expectations based on their IR luminosities and SFRs. 

We emphasize that, in term of their position in the $L_{\rm IR}-L'_{\rm CO}$ plane (cf. Fig. \ref{fig2}a), our sources and the lensed literature LBGs cB58 ($z=2.72$), the Cosmic Eye \citep[$z=3.07$,][]{riechers10}, and the MS1358-arc \citep[$z=4.93$,][]{livermore12} behave similarly. We caution though that the latter three are quite different sources due to their stellar masses which are  $\sim$2 orders of magnitude smaller than for customarily studied galaxies at $z<4$. Lensed LBGs have also much higher sSFR values ($\sim$25-60 Gyr$^{-1}$). Even accounting for a possibly  sublinear slope of the MS at $z=3$--4, they still display excesses of 4.2--6.0 above the MS, which would classify them as starbursting objects (Rodighiero~et~al.~2011). Nevertheless their CO-to-L$_{\rm IR}$ ratios are still factors of 2--3 smaller, even  than those of the most powerful local starbursting ULIRGs (Fig.2a, dashed line). Even if the properties of these lenses are harder to interpret, they hence appear to support evidence for weak CO emission compared to lower-z sources.

\subsection{Metallicity driven CO suppression}\label{lcoZ}

Figure~\ref{fig2}b re-displays the same galaxies of Fig.2a, showing their CO luminosity normalized by the average $L'_{\rm CO}$(SFR) trend 
of Fig.2a, and plotted as a function of redshift. Massive $z>3$ galaxies might depart from the trend of lower-z galaxies either because they have 
 enhanced SFR to gas mass ratios (i.e. high SFEs, implying departure from the integrated Schmidt-Kennicutt law -- S-K; SFR$\propto$M$_{\rm gas}^{1.2}$), or 
 due to high CO-to-H$_2$ conversion factors $\alpha_{\rm CO}$ caused by low metallicities. While we cannot solve the degeneracy with the
available evidence, we focus on the latter effect here based on recent developments in the determination of metallicity of $z>3$ galaxies \citep[e.g.,][]{maiolino08,sommariva12}.

To quantitatively explore the metallicity-driven suppression of CO-emission,  
we calculate simple models for the CO emission of galaxies based on their physical properties. We assume that normal galaxies at both $z\leq2.5$ and $z>3$ follow the same S-K law and have metallicity-dependent conversion factors. 
To a given $M_{\star}$ and redshift, we associate SFR- and $M_{\rm gas}$-values appropriate for an average main-sequence galaxy lying on the S-K law. From the $M_{\star}$ (and SFR)  we can infer the expected metallicity and thence $\alpha_{\rm CO}$ and our predicted $L'_{\rm CO}$ (using one of three approaches discussed below), which we normalize by the average $L'_{\rm CO}$(SFR) trend for MS galaxies from Fig.2a as for observed galaxies to display predictions in Fig.2b.
A rapid decrease of metallicity, as currently suggested by high-redshift measurements, would produce outliers with weak CO-emission. 

We use both the mass-metallicity ($M_{\star}-Z$) relation and the fundamental metallicity relation (FMR) relating metallicity to $M_{\star}$ and SFR to statistically assign gas-phase metallicities \citep{erb06,sommariva12,zahid13,mannucci10}. The metallicity-dependent $\alpha_{\rm CO}$ is assumed to scale as $\sim Z^{-1}$ (review by Bolatto et al. 2013; Sargent et al. 2013, in preparation, the latter study being based on the shape of the z=0 CO luminosity function \citep{keres03}). 
Fig.2b shows how these simple predictions accurately describe the location of galaxies at $0<z<2.5$. 
The smoothly varying metallicity for most galaxies over $0<z<2.5$ causes limited variations of $\alpha_{\rm CO}$, which results in a tight $L'_{\rm CO}$--SFR relation, and thus only weak scatter in both Fig.2 panels. However, the systematic decrease of $\alpha_{\rm CO}$ with increasing $M_{\star}$
produces small mass-dependent drifts that we correct in both data and models (see caption of Fig.2).

Figure~\ref{fig2}b shows tracks for the case of an FMR that remains constant at all redshift, for an FMR that breaks at z=2.5, and for a $M_{\star}-Z$ relation with accelerated evolution at $z>2.5$. Evidence for metallicity evolution at $z>2.5$ comes from \citet{mannucci10} and \citet{sommariva12}.
For the scenario with a broken FMR, we interpolate linearly with redshift the metallicity drop of 0.6 dex reported in \citet{mannucci10} between z=2.5 and z=3.3 and continue this trend beyond z=3.3 to estimate high-z evolution (but see \citealt{hunt12} for a non-evolutionary interpretation of the low-metallicity readings at high-redshifts).
For the case of an evolving $M_{\star}-Z$ relation, a similar linear interpolation/extrapolation is applied between z=2.2 and z=3 based on the $M_{\star}-Z$ relations of \citet{sommariva12} and \citet{zahid13}. A distinct difference among these predictions is the rapid decrease of the $L'_{\rm CO}/L'_{\rm CO}$(SFR) ratio at high-z in the case of fast metallicity evolution at z$>$2.5.  For the case of an universally valid FMR, the evolution of $L'_{\rm CO}/L'_{\rm CO}$(SFR) is not overly strong and almost vanishes at fixed $M_{\star}$. The scatter of the reference $z<2.5$ sample around the average predicted trend is 0.24 dex. 
M23 starts to deviate from the constant FMR track, agreeing better with the prediction assuming a rapid decline in metallicity at $z\sim3$. BD29079 displays an offset of $>2\sigma$ relative to the scenario of an universally-valid FMR.

\subsection{The CO to H$_2$ conversion factor in a regime of evolving metallicity} 

Metallicity evolution with redshift would affect, as discussed, the CO-to-H$_2$ conversion factor, making it challenging to infer the H$_2$-content from CO-measurements. The two massive LBGs studied here exemplify this: under the aforementioned, different plausible scenarios, 
$\alpha_{\rm CO}$ could range over 4--12 $M_{\odot}~(\rm{K~km~s}^{-1}\rm{pc}^2)^{-1}$ for M23 and 7--63 for BD29079. 
Inferring $\alpha_{\rm CO}$ at $z\gtrsim3$ is even much harder than at $z\sim1.5-2$. 
An overall better, average estimate of molecular gas mass could arguably be derived from the SFR itself, via the integrated S-K relation rather than from CO-emission (though we caution that no guarantee exists that the S-K still holds unchanged at $z>3$). Adopting the well-established relation between $M_{\rm gas}$ and SFR for normal galaxies as in S13a, we infer molecular gas masses $M_{\rm gas} \sim1.6\times10^{11}M_{\rm \odot}$ and  $M_{\rm gas} \sim1.8\times10^{11}M_{\rm \odot}$ for BD29079 and M23, respectively. With $M_{\rm gas}$=$\alpha_{\rm CO} L'_{\rm CO}$, the corresponding $\alpha_{\rm CO}$ would be $>$19 and $\gtrsim$9.3 for BD29079 and M23.

\subsection{The evolution of molecular gas fraction}
Recent studies reported a trend for increasing molecular gas fractions ($f_{\rm gas}=M_{\rm gas}/[M_{\rm gas}+M_{\star}]$) with redshift up to z$\sim$2.5 \citep{daddi10,geach11,magdis12a,tacconi13,saintonge13}, in good agreement with the redshift evolution of sSFR \citep{magdis12b}, suggesting that the peak epoch of cosmic star formation (z$\sim$2) corresponds to an epoch when typical massive galaxies were dominated by molecular gas. Some works have highlighted the impact of the assumed SFH and of nebular emission on SED-fitting and consequently on the measured sSFR-evolution beyond z$\sim$2 \citep{debarros13,gonzalez13}, in light of the disagreement between the observed sSFR plateau and current theoretical predictions \citep{neistein08,weinmann11}. Given the tight correlation between SFR and molecular gas content, the study of gas fractions in galaxies at z$>$3 might be useful to investigate the putative sSFR-plateau at z$\sim$2--8.

\begin{figure}[htbp]
\centering
\includegraphics[width=0.45\textwidth]{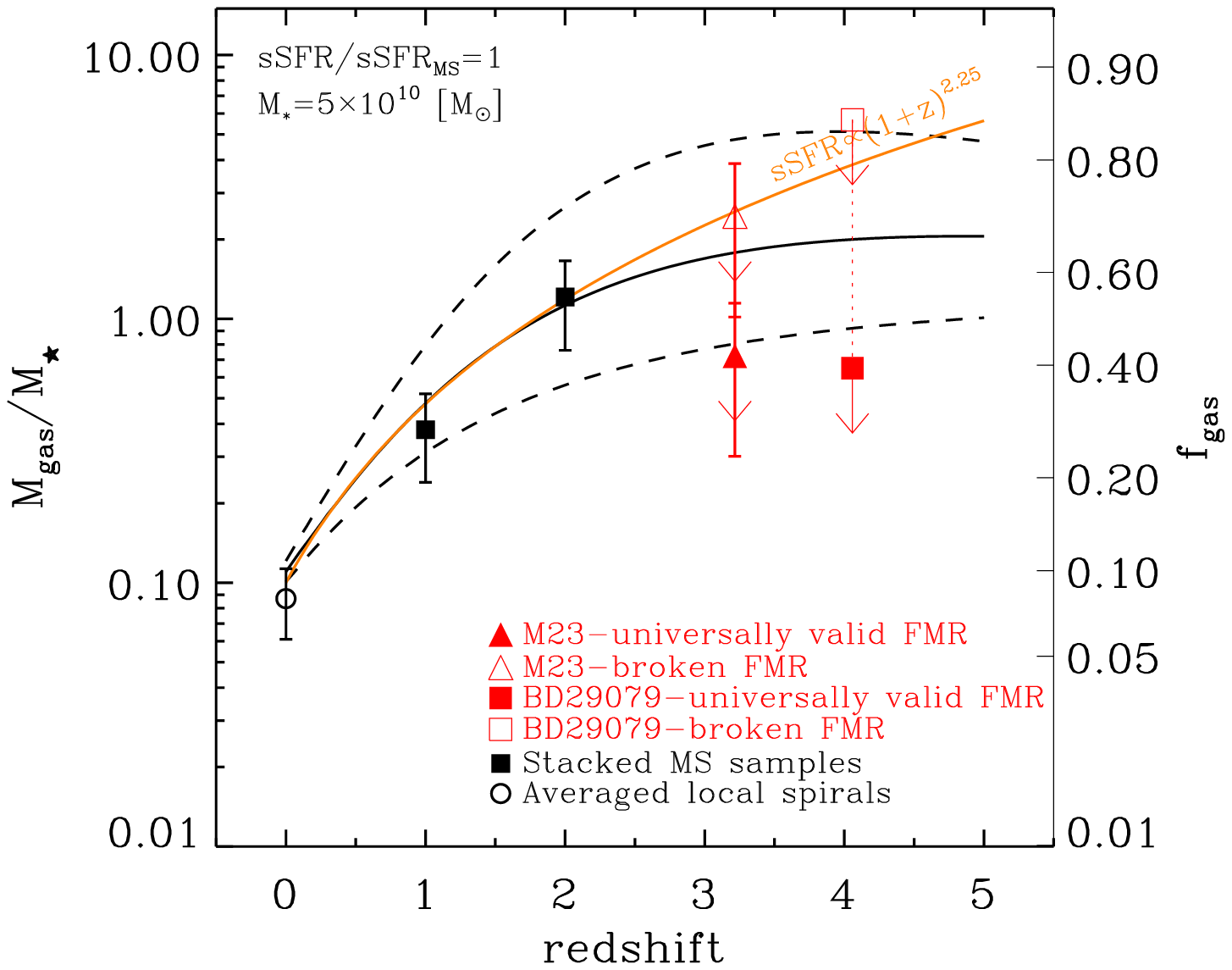}
\caption{\scriptsize Redshift-evolution of $M_{\rm gas}/M_{\star}$ and the molecular gas fraction of normal galaxies with sSFR/sSFR$_{\rm MS}$=1 and $M_{\star}=5\times10^{10}M_\odot$.  We compare the LBGs of this study to the $z=0$ sample of \citet{leroy08}, and stacked samples of $z\sim1$ and $z\sim2$ normal galaxies from \citet{magdis12b}. Gas fraction estimates for M23 and BD29079 derived assuming an universally valid (broken) FMR at $z\geq 3$ are marked with solid (open) red symbols. Black-solid line -- evolution of $M_{\rm gas}/M_{\star}$ at $M_{\star}=5\times10^{10}M_\odot$, predicted based on the observed sSFR-evolution (S13a; dashed line -- $\pm1\sigma$ uncertainties). The orange line traces the evolution predicted by the model of \cite{dave12}. \label{fig3}}
\end{figure}

Figure~\ref{fig3} shows the evolution of  $M_{\rm gas}/M_{\star}$ for normal galaxies with sSFR/sSFR$_{\rm MS}$=1 and $M_{\star}=5\times10^{10}\ M_\odot$  out to z$\sim$4. 
Gas fractions of galaxies with $M_{\star}\neq5\times10^{10}M_\odot$ were rescaled based on the observed trends $M_{\rm gas}/M_{\star}\propto{\rm (sSFR/sSFR_{MS}})^{0.9}$ and $M_{\rm gas}/M_{\star}\propto\ M_{\star}^{-0.5}$ \citep{magdis12b}. For the case of an universally valid FMR, we infer $M_{\rm gas}/M_{\star}$ of $\lesssim0.72$ and $<$0.65 for M23 and BD29079, which in both cases is lower than expected based on the observed sSFR-evolution, i.e., an increase with (1+z)$^{2.8}$ to z$\sim$2.5, followed by a plateau or a slow rise at even higher redshifts. Larger gas fractions 
are obviously recovered if, as in the previous sections, we allow for rapidly declining metallicities. 
If we assume strong metallicity-evolution at $z>2.5$, the gas fractions of M23 and BD29079 appear to agree with the measured sSFR-evolution, and also
with some theoretical predictions, e.g. for the case of a continuously increasing sSFR$\propto$(1+z)$^{2.25}$, as expected for accretion of pristine gas from the intergalactic medium \citep{dave12}. Given the low significance of our CO-detections and the large systematic uncertainties, our results can only delimit a range for the evolution of gas fractions at $z\sim\ 3$--4. Additional, direct metallicity estimates for high redshift galaxies would be highly beneficial for further exploration of the evolution of gas fractions. A larger sample of normal galaxies with CO-detections at $z>3$ is also clearly needed to reach firm conclusions.

\section{Conclusions}

We presented evidence that the CO-emission from two $z\gtrsim$3 massive LBG galaxies is weaker than expected based on the tight correlation between IR and CO luminosities satisfied by similar $z\leq2.5$ objects. A plausible explanation is that {\bf $z>3$} normal galaxies are weak in CO due to rapid metallicity-evolution at high redshift. As a result, the evolution of molecular gas fractions beyond $z\sim3$ is fairly uncertain due to the large uncertainties on metallicity-estimates and the shape of the relation between CO-to-H$_2$ conversion factor and metallicity. Also, the exact normalization and dispersion of the main-sequence is not well constrained at these redshifts, and their attenuation properties might have evolved, thus making it even harder to distinguish normal disk-like galaxies from merger-powered sources. Our observations (similarly to results of \citealt{ouchi13}) in any case start to point to the difficulty of detecting gas cooling emission lines from normal galaxies at $z>3$. This might remain challenging even with the unprecedented sensitivity of ALMA if metallicities indeed decrease rapidly for typical massive galaxies at these high redshifts.

\acknowledgments
This work was based on observations carried out with the IRAM PdBI, supported by INSU/CNRS (France), MPG (Germany), and IGN (Spain). We thank C. Feruglio and M. Krips for help during observations/data reduction, and the anonymous referee for valuable comments. QT, ED, MTS and MB acknowledge funding from ERC-StG grant UPGAL 240039 and ANR-08-JCJC-0008.

\end{document}